# Magnetic scattering and electron pair breaking by rare-earth-ion substitution in $BaFe_2As_2$ epitaxial films

Takayoshi Katase[1,4], Hidenori Hiramatsu[2,3], Toshio Kamiya[2,3] and Hideo Hosono[1,2,3,5]

[1] Frontier Research Center, Tokyo Institute of Technology, S2-6F East, 4259 Nagatsuta-cho, Midori-ku, Yokohama 226-8503, Japan

[2] Materials and Structures Laboratory, Tokyo Institute of Technology, 4259 Nagatsuta-cho, Midori-ku, Yokohama 226-8503, Japan

[3] Materials Research Center for Element Strategy, Tokyo Institute of Technology, 4259 Nagatsuta-cho, Midori-ku, Yokohama 226-8503, Japan

[4] Present address: Research Institute for Electronic Science, Hokkaido University, Sapporo 001-0020, Japan

[5] Corresponding author. E-mail: hosono@msl.titech.ac.jp

**Abstract.**

The effect of electron doping by trivalent charge state rare-earth ion (*RE* = La, Ce, Pr, and Nd) substitutions on the superconductivity in $BaFe_2As_2$ was examined using epitaxial films. Each of the *RE* substitutions suppressed the resistivity anomaly associated with the magnetic/structural phase transitions, leading to the resistivity drops and superconductivity transitions. Bulk superconductivity was observed at the maximum onset critical temperature ($T_c^{onset}$) of 22.4 K for La-doping and 13.4 K for Ce-doping, while only broad resistivity drops were observed at 6.2 K for Pr-doping and 5.8 K for Nd-doping but neither zero resistivity nor distinct Meissner effect were observed at least down to 2 K. The decrease in $T_c^{onset}$ with increasing the number of *RE* 4*f* electrons cannot be explained in terms of the crystalline qualities or crystallographic structure parameters of the $BaFe_2As_2$ films. It was clarified, based on resistivity-temperature analyses, that magnetic scattering became increasingly significant in the above order of the *RE* dopants. The negative magnetoresistance was enhanced by the Ce- and Pr-doping, implying that the decrease in $T_c$ originates from magnetic pair breaking by interaction of the localized 4*f* orbitals in the *RE* dopants with the itinerant Fe 3*d* orbitals.

## 1. Introduction

In the few months since the report on high critical temperature ($T_c$) superconductivity at 26 K in the 1111-type iron pnictide $LaFeAs(O_{1-x}F_x)$ [1], 122-type $AEFe_2As_2$ (where $AE$ = alkaline earth) [2] has also joined the family of iron-based high-$T_c$ superconductors as a parent material. To induce its high-$T_c$ superconductivity, both types of doping carrier, i.e. holes and electrons, are typically used by selection of appropriate aliovalent dopants. To date, most such carrier doping processes have been performed by substituting the *AE* sites with alkali metals with different ion charges, such as K (e.g., in hole-doped $(Ba_{1-x}K_x)Fe_2As_2$) [2], and by substituting the Fe sites with transition metals with excess 3*d* electron numbers, such as Co (e.g., in electron-doped $Ba(Fe_{1-x}Co_x)_2As_2$) [3]. The doping sites are categorized into two modes for 122-type $AEFe_2As_2$, i.e., “indirect doping” for doping at sites other than the Fe sites, and “direct doping” for doping at the Fe sites, because the *AE* and the FeAs layers are spatially separated and the Fermi level is composed mainly of Fe 3*d* orbitals [4].

The maximum $T_c$ value for each parent material of the iron-based superconductors was obtained by indirect doping [2],[5–7] because direct doping has a major influence on carrier transport in the conducting Fe layer. It was therefore expected that a new indirect electron doping mode for $AEFe_2As_2$, e.g. at the *AE* sites, would lead to high-$T_c$ superconductors, similar to the effects of other indirect doping methods, such as K doping at *AE* sites (maximum $T_c$ = 38 K) [2] and isovalent P doping at As sites (maximum $T_c$ = 31 K) [8]. However, indirect electron doping of $AEFe_2As_2$ by substituting the *AE* sites with trivalent rare earth (*RE*) ions was difficult to perform by conventional solid-state reactions; only Ba [9] and Sr [10] have been reported. This difficulty was attributed to electronic instability arising from the high localized density of states at the Fermi level, as predicted for La-doped $AEFe_2As_2$ [$(AE_{1-x}La_x)Fe_2As_2$] by density functional theory calculations [11,12].

Under such circumstances, indirect $RE$ doping of $AEFe_2As_2$ was achieved by applying a high-pressure synthesis process for $(Sr_{1-x}La_x)Fe_2As_2$ polycrystals [10] and a melt-growth process using a flux agent for $(Ca_{1-x}RE_x)Fe_2As_2$ (where $RE$ = La – Nd) single crystals [13−16]. Among these materials, it should be noted that a Pr-doped $CaFe_2As_2$ single crystal demonstrated a maximum $T_c$ of 49 K [14], which is the highest reported $T_c$ among the 122-type $AEFe_2As_2$ series, although their shielding volume fractions are as low as <1 % at 40 K [13].

In contrast to the $AE$ = Ca and Sr systems, there had been no reports on $RE$-doped $BaFe_2As_2$, probably because of the large difference between the ion radii of $Ba^{2+}$ (142 pm for 8-fold coordination) and $RE^{3+}$ (111−116 pm) [17]. However, we recently succeeded in indirect La doping at the Ba sites in $BaFe_2As_2$ by using a film-growth process [18]. This success was attained by using the highly non-equilibrium nature of thin film deposition via the vapor phase. The maximum $T_c$ (22.4 K) of the $(Ba_{1-x}La_x)Fe_2As_2$ films was much lower than that of the $RE$-doped $CaFe_2As_2$ single crystal, and its electronic phase diagram was almost the same as that of directly-doped $Ba(Fe_{1-x}Co_x)_2As_2$. On the other hand, according to the scenario in $(Ca_{1-x}RE_x)Fe_2As_2$ crystals where a smaller $RE$ dopant than La, such as Pr, leads to a higher $T_c$, we expected that substitution of smaller $RE$ ions with open-shell 4*f* electrons for the Ba sites would provide a higher $T_c$ than that attained for $(Ba_{1-x}La_x)Fe_2As_2$.

In this paper, we examined the indirect electron doping of open 4*f* shell $RE$ ions (Ce, Pr, Nd, and Sm) at the Ba sites in $BaFe_2As_2$ epitaxial films by pulsed laser deposition (PLD). It was confirmed that three open 4*f* shell $RE$ dopants (Ce, Pr, and Nd) were successfully substituted at the Ba sites by the PLD process, similar to the case for La doping. The Ce doping induced bulk superconductivity at a maximum onset $T_c$ ($T_c^{onset}$) of 13.4 K. On the other hand, although zero resistivity and distinct Meissner effect were not observed down to 2 K, resistivity drops caused by superconducting transitions were observed at 6.2 K for the Pr doping and 5.8 K for the Nd doping. The relationship between the 4*f* electron configuration and the superconducting properties is discussed and comparison is made with the results of the non-magnetic La doping case.

## 2. Experimental

$(Ba_{1-x}RE_x)Fe_2As_2$ thin films were grown directly on MgO (001) single crystals with a PLD system using a second harmonic of a Nd:YAG laser as the excitation source [19]. Bulk polycrystalline samples of *RE* (La, Ce, Pr, Nd, and Sm)-containing $BaFe_2As_2$ were used as PLD targets (see supplementary Fig. S1 for the synthesis of the bulk polycrystals). Powder X-ray diffraction (XRD, using CuKα anode radiation) analyses confirmed that the crystalline phases of the bulk polycrystals were mixtures of $BaFe_2As_2$ and *RE*As impurity phases, i.e. the *RE* ions were not incorporated in the $BaFe_2As_2$ phases [18]. Film growth was carried out at the previously-optimized substrate temperature ($T_s$) of 850°C (note that the *RE*As impurity phases appeared in the films at higher values of $T_s$ > 850°C) [18–20]. Each film was 150–250 nm thick, as measured with a stylus profiler. The net doping concentrations in the $(Ba_{1-x}RE_x)Fe_2As_2$ films ($x_{film}$) were measured using an electron-probe microanalyzer (EPMA). The high homogeneity of the *RE* distribution was confirmed by the EPMA for all of the fabricated samples (see supplementary Fig. S2(a) for mapping images of the *RE* concentrations). It was also confirmed that $x_{film}$ was controlled by changes in the nominal *RE* concentration (nominal $x$) of the *RE*-containing $BaFe_2As_2$ PLD targets (see supplementary Fig. S2(b) for the relationship between nominal $x$ and $x_{film}$). The film structures and their crystalline qualities were characterized by high-resolution XRD using Cu K$\alpha_1$ anode radiation at room temperature. The temperature dependence of the electrical resistivity ($\rho$–$T$) was measured by the four-probe method in a temperature range of 2–300 K with a physical property measurement system. Temperature dependence of the magnetic susceptibility ($\chi$–$T$) was measured with a vibrating sample magnetometer after zero-field cooling (ZFC) and during field cooling (FC).

**3. Results**

A. Growth of $(Ba_{1-x}RE_x)Fe_2As_2$ thin films

Figure 1 shows the out-of-plane XRD patterns for films of (a) $(Ba_{1-x}Ce_x)Fe_2As_2$, (b) $(Ba_{1-x}Pr_x)Fe_2As_2$, (c) $(Ba_{1-x}Nd_x)Fe_2As_2$, and (d) Sm-containing $BaFe_2As_2$ with various values of $x_{film}$. In the cases of (a) Ce, (b) Pr, and (c) Nd doping, the $(Ba_{1-x}RE_x)Fe_2As_2$ films obtained were grown epitaxially on MgO substrates with the epitaxial relationship of [001] $(Ba_{1-x}RE_x)Fe_2As_2$ || [001] MgO for the out-of-plane case and [100] $(Ba_{1-x}RE_x)Fe_2As_2$ || [100] MgO for the in-plane case, which is same as that of $(Ba_{1-x}La_x)Fe_2As_2$ films [18]. The 00*l* diffraction angles of the undoped $BaFe_2As_2$ film are indicated by the dotted lines to clearly show the peak shift caused by the *RE* substitutions. Sharp 00*l* diffractions of the $(Ba_{1-x}RE_x)Fe_2As_2$ phases, along with those from a small amount of the Fe impurity phase (indicated by the asterisks), were observed. The 00*l* diffraction peaks shifted systematically to higher angles from the diffraction peaks of the undoped $BaFe_2As_2$ film as $x_{film}$ increased. The intensities of the Fe impurity diffractions were almost the same in all the samples. Segregation of the impurity phases of *RE*As started to be observed at the high $x_{\mathrm{film}}$ values of 0.30, 0.28, and 0.20 for the Ce-, Pr-, and Nd-doped films, respectively, from which we determined the solubility limits to be lower than these values. Because the formation enthalpies of the *RE*As impurity phases are almost the same (−288 kJ/mol for CeAs, −307 kJ/mol for PrAs, and −304 kJ/mol for NdAs) [22], the different solubility limits are attributed to the differences in the ion size mismatch between $Ba^{2+}$ and $RE^{3+}$ in each case, and in the consequent instability among the $(Ba_{1-x}RE_x)Fe_2As_2$ phases. In contrast, in the case of Sm doping with $x_{\mathrm{film}}$ = 0.1 and 0.17, no peak-shift of the 00*l* diffractions was detected, and only the segregation of SmAs impurities was observed (Fig. 1(d)), indicating that the incorporation of the smaller Sm ions into $BaFe_2As_2$ was unsuccessful.

B. Structural characterization

Figure 2(a) summarizes the evolution of the lattice parameters *a* and *c* and the unit cell volume *V* at room temperature as a function of $x_{\mathrm{film}}$ for $(Ba_{1-x}RE_x)Fe_2As_2$ (*RE* = Ce, Pr, and Nd) epitaxial films. Note that those of the films having the segregated *RE*As phases in Fig. 1 are not plotted because the $x_{\mathrm{film}}$ exceeds the solubility limits and the lattice parameters remained almost unchanged from those of the highest $x_{\mathrm{film}}$ films in Fig. 2. The *c*-axis and *a*-axis lattice parameters were determined by out-of-plane and in-plane XRD, respectively. In all of the dopant cases, systematic shrinkages of the *c*-axis length were observed (the largest $\Delta c/c$ was ~ –2.3% for $(Ba_{1-x}Ce_x)Fe_2As_2$ films with $x_{\mathrm{film}}$ = 0.28), while the shrinkage of the *a*-axis length was very small (the largest $\Delta a/a$ was ~ –0.3%). Consequently, *V* decreases monotonically as $x_{\mathrm{film}}$ increases. These results, which are similar to those obtained for $(Ba_{1-x}La_x)Fe_2As_2$ [18], substantiate the fact that the $RE^{3+}$ ions substitute the $Ba^{2+}$ sites in the epitaxial films. It is noteworthy that the *c*-axis shrinkage increases in the order of Nd, Pr, and Ce dopants, when compared with the same $x_{\mathrm{film}}$, but this result is inconsistent with the differences in the ion radii of the $RE^{3+}$ ions (their radii decrease in the order of $Ce^{3+}$ (114 pm), $Pr^{3+}$ (113 pm), and $Nd^{3+}$ (111 pm) because of the lanthanide contraction) [17]. Additionally, the variation in the *a*-axis length was independent of the ion radii.

Figure 2(b) shows the $x_{\mathrm{film}}$ dependence of the crystalline quality of the $(Ba_{1-x}RE_x)Fe_2As_2$ epitaxial films. Here, the full width at half maximum (FWHM) values of rocking curves for 004 ($\Delta\omega$, out-of-plane) and 200 ($\Delta\phi$, in plane) diffractions were used to evaluate the crystalline quality. The $\Delta\omega$ and $\Delta\phi$ values scatter to an extent but remained almost unchanged at ~1.0 degree, regardless of $x_{\mathrm{film}}$ for all dopants. Figure 2(c) shows the $x_{\mathrm{film}}$ dependence of the FWHM values of 004 diffractions ($\Delta 2\theta$). $\Delta 2\theta$ gradually increased as $x_{\mathrm{film}}$ increased, which may originate from the structural strains and/or distortions in the films, probably because of large ion-size mismatches between the $Ba^{2+}$ ion and the doped $RE^{3+}$ ions. However, it is safely concluded that the structural quality is similar for all *RE* dopants.

C. Transport and magnetic properties

Figure 3 summarizes the $\rho$–$T$ curves for the epitaxial films of (a) $(Ba_{1-x}Ce_x)Fe_2As_2$, (b) $(Ba_{1-x}Pr_x)Fe_2As_2$, and (c) $(Ba_{1-x}Nd_x)Fe_2As_2$ for various values of $x_{\mathrm{film}}$. The inset figures are magnified views which show the resistivity drops more clearly. The $\rho$–$T$ curve of an undoped $BaFe_2As_2$ epitaxial film is shown in the top panel of Fig. 3(a) for comparison. The $\rho$ of an undoped $BaFe_2As_2$ film decreases with decreasing $T$ from 300 K, and falls rapidly from ~150 K, whose resistivity anomaly is associated with magnetic/structural transitions [23]. As seen in supplementary Fig. S3, a $\mathrm{d}\rho/\mathrm{d}T$ – $T$ curve provides a clear peak and resistivity anomaly temperatures ($T_{anom}$). It should be noted that the anomalous temperature range around $T_{anom}$ for the undoped $BaFe_2As_2$ epitaxial film is broader than that of a single crystal [24]. However, the crystalline quality of this film ($\Delta\omega$ of the 002 diffraction ~ 1 deg.) is almost the same as that of the single crystal ($\Delta\omega$ of the 002 diffraction = 0.7 deg.) [24]. Further, it is reported that a sharp $d\rho/dT$ curve similar to that of the single crystal is observed even in a polycrystal [25]. These results indicate that the broader magneto-structural transition of this film does not originate from a crystalline quality issue. It is reported that a small in-plane stress applied to the $BaFe_2As_2$ single crystal broadens the structural transition due to the de-twinning of the crystals [26]; therefore, we speculate that a lattice-strain effect at the epitaxial film – substrate interface would be an origin of the broadening. In all dopant cases, $\rho$ at 300 K gradually decreased, and $T_{anom}$ shifted to lower temperatures as $x_{film}$ increased.

In the Ce-doping case (a), a resistivity drop without zero resistivity was observed for $x_{film}$ = 0.09, but the resistivity anomaly was still observed at $T_{anom}$ = 70 K. With a further increase in $x_{film}$, the resistivity anomaly was not detected in the $\rho$–$T$ curves, and $T_c^{onset}$ for a superconductivity transition appears at $x_{film} \geq 0.09$. $T_c^{onset}$ reached a maximum value of 13.4 K at $x_{film}$ = 0.15. The resistivity transition width (defined by $\Delta T_c = T_c^{onset} - T_c^{offset}$, where $T_c^{offset}$ is the offset critical temperature determined by extrapolating a $\rho$–$T$ curve to zero resistivity) of this film is 4.5 K, which is doubly larger than $\Delta T_c$ = 2.7 K of La-doped films [18] although the crystalline qualities are almost the same as seen in $\Delta\phi$, $\Delta\omega$ and $\Delta 2\theta$ in Figs. 2(b,c), and the dopant distribution is homogeneous, which is confirmed both by EPMA (see supplementary Fig. S2(a)) and XRD (peak shift and broadening are not observed by Ce doping). Thus, the broad resistivity transition reflects a wide vortex-liquid-phase due to strong vortex pinning centers, similar to that of the Co-doped $BaFe_2As_2$ epitaxial film [27,28]. The wide liquid phase is due to disorders in the films, which may also the case in the Ce-doped films. $T_c^{onset}$ then decreased as $x_{film}$ increased further, and the resistivity drop finally disappeared at $x_{film}$ = 0.29. Also, in the case of Pr doping (b), $T_{anom}$ shifted to lower $T$ values as $x_{film}$ increased and the resistivity drop was observed at $x_{film}$ = 0.11. However, although $T_{anom}$ disappeared at $x_{film}$ = 0.18, clear zero resistivity was not observed at least down to 2 K. On the other hand, in the Nd doping case (c), the resistivity drop was observed at $x_{film}$ = 0.13, but larger values of $x_{film}$, where the resistivity anomaly is completely suppressed, could not be obtained because of the low solubility limit of the Nd dopant. The maximum transition temperatures of the resistivity drop are 6.2 K and 5.8 K for Pr- and Nd-doping, respectively, but each film did not get into clear superconducting state.

To confirm that the observed resistivity drops originate from superconducting transitions, we measured the magnetic field dependence of the $\rho$–$T$ curves and the temperature dependence of the magnetic susceptibilities ($\chi$–$T$) for epitaxial films of $(Ba_{0.85}Ce_{0.15})Fe_2As_2$ (a), $(Ba_{0.82}Pr_{0.18})Fe_2As_2$ (b), and $(Ba_{0.87}Nd_{0.13})Fe_2As_2$ (c), whose chemical compositions were chosen to have the maximum $T_c^{onset}$ for each dopant (Fig. 4). For both measurements, the external magnetic field was applied parallel to the *c*-axis of the epitaxial films. For all dopants, shifts of $T_c$ to lower $T$ were observed under application of magnetic fields. This result strongly suggests that all the low-temperature resistivity drops originate from superconducting transitions. A clear diamagnetic signal with a shielding volume fraction (SVF) of up to 10% at < 4 K was observed only in the case of Ce doping. Note that, the Meissner effect starts to appear at a temperature of $T_c^{Meissner}$ ~ 7.8 K, which is almost the same as $T_c^{zero}$ ~ 8.0 K where the resistivity reaches zero. However, the SVFs of the superconducting phases in the Pr- and Nd-doped films are very low because no Meissner effect is observed in their $\chi$–$T$ curves. This observation suggests a possibility that the observed resistivity drops in the Pr- and Nd-doped films come from a superconducting transition from impurity phases; however, the impurity phase detected by XRD (Fig. 1) was only Fe, which does not show superconductivity at $T \geq 2$ K. It is hard to consider that an amorphous impurity caused the superconductivity-like resistivity drops; therefore, we consider that the Meissner effect was not observed for the Pr- or Nd-doped film because $T_c^{Meissner}$ would be lower than the temperature range examined in this study (i.e., < 2 K). As a consequence, we conclude that the resistivity drops are caused by superconducting transitions of the Pr- and Nd-doped $BaFe_2As_2$ phases. Considering that perfect diamagnetism with an SVF of 100% at $T < 7$ K was obtained in the case of non-magnetic La-doped $(Ba_{1-x}La_x)Fe_2As_2$ epitaxial films with maximum $T_c^{onset}$ = 22.4 K [18], it seems that the doping of *RE* with 4*f* electrons decreases the SVF of its superconducting phases in the extreme. This result is markedly different from that in the case of the $(Ca_{1-x}RE_x)Fe_2As_2$ crystals because the maximum $T_c^{onset}$ of the *RE*-doped $CaFe_2As_2$ single crystal increases in the order of *RE* = La, Ce, Pr and Nd, which is opposite to the present $(Ba_{1-x}RE_x)Fe_2As_2$ case [13–16].

D. Electronic phase diagrams

Figure 5 summarizes the electronic phase diagrams for the $(Ba_{1-x}RE_x)Fe_2As_2$ epitaxial films. The electronic phase diagram of $(Ba_{1-x}La_x)Fe_2As_2$ epitaxial films is also shown in the figure for comparison [18]. For all *RE* dopants, the antiferromagnetic (AFM) transition at $T_{anom}$ is suppressed as $x_{film}$ increases, but the suppression rate of $T_{anom}$ is different for different dopants; i.e., it decreases in the order of *RE* = La, Ce, Pr, and Nd. The vertical arrows, which indicate the extrapolated intersections at the $x_{film}$ axis by using a phenomenological fit to $T_{anom}$, indicate this trend more clearly. The extrapolated $x$-point (critical $x$), where the resistivity anomaly would disappear completely, shifts to higher $x_{film}$ values in the order of La-, Ce-, and Pr-doping (in the case of Nd doping, the extrapolation is unavailable because of the low solubility limit).

The plots of $T_c^{onset}$ for the $(Ba_{1-x}Ce_x)Fe_2As_2$ epitaxial films form a dome-shaped structure in the $0.06 \leq x_{film} \leq 0.28$ range. Also, in the cases of Pr- and Nd-doping, $T_c$ appeared at $x_{film} > 0.1$ as $T_{anom}$ shifted to the lower $T$ side. However, unlike the Ce-doping case, neither of the superconducting domes extended to an over-doped region. The maximum $T_c^{onset}$ values for the $(Ba_{1-x}Ce_x)Fe_2As_2$ and $(Ba_{1-x}Pr_x)Fe_2As_2$ films were obtained at their critical $x$-points of ~0.15 and ~0.18, respectively. The coexistence of superconductivity with an AFM state is stabilized only in the doping region that is lower than the critical $x$. The superconducting dome shape and the coexistence of AFM with superconductivity are qualitatively similar to those observed in the $(Ba_{1-x}La_x)Fe_2As_2$ epitaxial films, but the superconducting dome is slightly shifted to a higher $x_{film}$ values and the height of the dome becomes smaller than that of $(Ba_{1-x}La_x)Fe_2As_2$, suggesting that the maximum $T_c^{onset}$ at the critical $x$ point decreases depending on the *RE* dopant species. The difference in the suppression rate of $T_{anom}$ and the superconducting dome shift may relate to the different variations in the $c$-axis lattice parameters with respect to $x_{film}$, as shown in Fig. 2(a). The inset figure in Fig. 5 replots $T_{anom}$ and $T_c^{onset}$ against the $c$-axis lattice parameters for the $(Ba_{1-x}RE_x)Fe_2As_2$ epitaxial films. The suppression curves of $T_{anom}$ converge to a single line, and the peak position of the superconducting dome converges to the same *c*-axis length of 1.285 nm. This result implies that both chemical pressure and carrier doping are effective for suppression of the AFM transition and induction of superconductivity in the $BaFe_2As_2$ system [29].

## 4. Discussion

First, we should remind that the amounts of the Fe impurity in the $(Ba_{1-x}RE_x)Fe_2As_2$ films were almost the same in all the samples with the different *RE* dopants; therefore, the severe decrease in $T_c$ and superconducting volume fractions cannot be explained by the impurity effects such as Fe. There are other plausible origins of the differences in the maximum $T_c^{onset}$ and SVFs of the superconducting phase of the $(Ba_{1-x}RE_x)Fe_2As_2$ films with the different *RE* dopants. The first is the structural differences, because it has been noted that the shapes and sizes of the $FeAs_4$ tetrahedra critically affect the superconducting properties of iron pnictide superconductors; i.e. deviation of the As–Fe–As bonding angle from a regular tetrahedron is highly correlated with the maximum $T_c$ values achieved [30]. Theoretical works have also shown that the Fermi surface topology is highly sensitive to the height of the As atom from the Fe sheet [31]. $BaFe_2As_2$ has a tetragonal crystal structure with a space group of *I*4/*mmm*, in which only the *z*-coordinate of the As ($z_{As}$, at the Wyckoff position 4*e*) site is a variable parameter. The height of the As atom from the Fe sheet ($h_{As} = c(z_{As}-0.25)$) and the As-Fe-As angle ($\alpha = 2\tan^{-1}(a/2h_{As})$) can therefore be estimated if the *a*-axis and *c*-axis lattice parameters and $z_{As}$ are determined. It has been reported that $z_{As}$ can be estimated from the ratio of the integrated intensities of 00*l* diffractions of epitaxial films in the out-of-plane XRD patterns [32]. In the Ce-doping case, it was found that $z_{As}$ only changed from $z_{As} = 0.3545$ at $x = 0$ to 0.352, even when $x_{film}$=0.29, because the intensity ratios of the 00*l* diffractions showed little change, as shown in Fig. 1(a), which indicates that the angle α and $h_{As}$ primarily depend on the lattice parameters [33]. Table I summarizes the superconducting and structural parameters of the following optimally doped epitaxial films with critical *x*: $(Ba_{0.87}La_{0.13})Fe_2As_2$ with maximum $T_c^{onset}$ of 22.4 K, $(Ba_{0.85}Ce_{0.15})Fe_2As_2$ with $T_c^{onset}$ of 13.4 K, and $(Ba_{0.82}Pr_{0.18})Fe_2As_2$ with $T_c^{onset}$ of 6.2 K. The lattice parameters and the structural parameters, such as $z_{As}$, $h_{As}$, and the α angles, are almost the same for each dopant. Density functional structure relaxation calculations for the models of $(Ba_{3/4}RE_{1/4})Fe_2As_2$ supported this result in that there is no remarkable difference in $z_{As}$, $h_{As}$, and the α angles for different *RE* dopants. These results indicate that the difference in the maximum $T_c$ and the SVFs of the superconducting phases cannot be explained in terms of difference in the local structure.

Another plausible origin is the different electronic properties that originate from the different numbers of the *RE* 4*f* electrons. It should be noted that the $\rho$–$T$ curves in Fig. 3 have a minimum at $T_{min}$ (shown by closed triangles) and that the resistivities show an upturn in the *T* region lower than $T_{min}$ for the $(Ba_{1-x}RE_x)Fe_2As_2$ epitaxial films with $x_{film}$ larger than the critical *x*. We used conventional power-law behavior, where $\rho = \rho_0 + AT^n$ (where $\rho_0$ is the residual resistivity), for these $\rho$–$T$ curves at $T>T_{min}$ to evaluate the electron scattering. The fitting results, indicated by the black lines in Fig. 3, agree well with the experimental $\rho$–$T$ curves. The resistivity upturn can be explained either by carrier localization with disorder scattering [34] or by magnetic scattering [35]. Therefore, the origin of the weak resistivity-upturn behavior for the non-magnetic La-doping should lie in carrier localization induced by defects and/or disorders generated by the large difference in ion radii between $Ba^{2+}$ and $La^{3+}$, which causes local structural distortions. Because the resistivity upturn at $<T_{min}$ is very weak in the case of the non-magnetic La dopant, this indicates that the extra resistivity upturns for *RE* = Ce and Pr originate from magnetic scattering caused by the magnetic *RE* dopants. Therefore, a possible mechanism for breaking of the superconducting electron-pair is the interaction between the localized *RE* 4*f* electrons and the Fe 3*d* conduction electrons [36]. The contribution of the magnetic *RE* ions to the electron scattering can be estimated by comparing the electronic properties of the *RE*-doped samples with those of the nonmagnetic reference $(Ba_{1-x}La_x)Fe_2As_2$ epitaxial films, because their structural differences are negligible, as shown in Fig. 2 and Table I.

The excess resistivity caused by a magnetic scattering should give rise to a negative magnetoresistance (MR), because a high magnetic field suppresses spin inversion and scattering [35]. To clarify the magnetic scattering behavior, the field dependence of MR, which is defined as $[\rho(\mu_0 H)-\rho(0)]/\rho(0)\times 100$, was measured at 30 K under fields of up to $\mu_0 H$ = 9 T applied parallel to the *c*-axis and at the DC-current in the *ab*-plane of the epitaxial film. Figure 6(a–c) shows the MR for epitaxial films of $(Ba_{1-x}La_x)Fe_2As_2$ with $x_{\mathrm{film}}$ = 0–0.44, $(Ba_{1-x}Ce_x)Fe_2As_2$ with $x_{\mathrm{film}}$ = 0.06–0.29, and $(Ba_{1-x}Pr_x)Fe_2As_2$ with $x_{\mathrm{film}}$ = 0.06–0.18. No asymmetry against negative fields and hysteresis were observed in the cyclic measurements. The MR of the undoped $BaFe_2As_2$ epitaxial films, shown in Fig. 6(a), showed a large positive MR up to +15% at $\mu_0 H$ = 9 T. Similar to the results of a previous report on undoped $BaFe_2As_2$ single crystals [37, 38], the MR of the undoped $BaFe_2As_2$ epitaxial film does not follow the standard Kohler's rule with $H^2$ dependence, and the crossover from the $H^2$ dependence at lower fields to the linear dependence at higher fields is observed at $\mu_0 H$ = 6.3 T. The observed MR can be fitted well by a modified two-band galvanomagnetic model ($MR_{\mathrm{SDW}}$) [39], in which the positive MR is caused by a magnetically ordered state that originates from spin-density-wave (SDW) gaps at low temperatures <$T_{anom}$. The $MR_{SDW}$ is expressed by $\Delta\rho/\rho_0 = 4(\beta-\gamma H)\mu_0^2H^2[(1+\beta-\gamma H)^2+(1-\beta+\gamma H)^2\,\mu_0^2H^2]^{-1}$, where $\beta$ is the ratio of the itinerant majority carrier concentration to the minority carrier concentration at $\mu_0 H = 0$ T, $\gamma$ is a constant value representing the magnetic field effect on the ratio of carrier concentration, and $\mu_0$ is the mobility of majority carriers, respectively. In the case of the non-magnetic La doping, the positive MR decreases with the increase in $x_{film}$ and disappears at the critical $x$ of 0.13. All of the positive MR observed on the La-doped films is explained very well by the $MR_{SDW}$ model, as shown by the fitting curves in Fig. 6(a). The fitted parameters are summarized in Table 2. The $\beta$ values increase and $\mu_0$ decrease with increasing the $x_{Film}$, while $\gamma$ values remain constant, which is consistent with the La-doping result that the La-doping increased the majority carriers (electrons) and decreased their mobilities. The MR became undetectable at higher $x_{film}$ values of 0.13–0.44, which is similar to the behavior of $Ba(Fe_{1-x}Co_x)_2As_2$ single crystals [40]. On the other hand, in the cases of magnetic Ce- and Pr-doping, a negative MR appeared as $x_{film}$ increased. It has been reported that MR with magnetic scattering represents a $-H^2$ dependence at low fields, and it then follows a −log $H$ dependence after sufficient suppression of the spin inversion [41,42]. It is seen that carrier doping into the $BaFe_2As_2$ films by the non-magnetic La dopants generate $MR_{SDW}$ only, while negative MR expressed by $-H^2$ and −log $H$ are superimposed for the magnetic Ce- and Pr-doping cases. The fitting results are shown by the curves in Figs. 6(b) and (c). The coexistence of a positive MR and a weak negative MR with $-H^2$ dependence was observed for low $x_{film} < 0.12$. With an increase in $x_{film}$, the positive MR disappeared and only the negative MR was then observed. In the case of Ce-doping with $x_{film}$ up to 0.18, the negative MR showed a $-H^2$ dependence at fields up to 9 T. The MR at larger values of $x_{film}$ up to 0.29 deviates from the $-H^2$ dependence at a higher value of $\mu_0 H$ of 3 T and showed a transition to the –log $H$ dependence. With decrease in temperature from 30 K to 2 K, the negative MR becomes larger and the transition $H$ value from the $-H^2$ dependence to the –log $H$ dependence shifts to lower fields (see supplementary Fig. S4 for MR measured at 30, 10, and 2 K for $(Ba_{0.72}Ce_{0.28})Fe_2As_2$ epitaxial film). In addition, the negative MR along the *ab*-plane is much smaller than that along the *c*-axis. This result indicates that antiferromagnetic spin-ordering parallel to the *c*-axis and ferromagnetic ordering parallel to the *ab*-plane affect the negative MR (see supplementary Fig. S4). Also, in the Pr-doping case, the field dependence changed from $-H^2$ to –log $H$ in the negative MR observed at a high $\mu_0 H$ of 2.5 T, even at a low $x_{\rm film}$ value of 0.18. With an increase in the numbers of 4*f* electrons from the *RE* dopants, the negative MR increased (the observed maximum negative MR = –0.15% for the Ce doping and –0.42% for the Pr doping with the same $x_{\rm film}$ value of 0.18.) No observation of the negative MR in the $(Ba_{1-x}La_x)Fe_2As_2$ films and the enhanced negative MR in the Ce- and Pr-doping cases, which is associated with spin flip magnetic scattering of the charge carriers, indicate that the interaction between the 4*f* electrons and the conduction 3*d* electrons is enhanced in the order from Ce- to Pr-doping. This would cause the electron pair-breaking that is responsible for the suppression of superconductivity in the magnetic *RE*-ion doped $(Ba_{1-x}RE_x)Fe_2As_2$ epitaxial films.

As noted at the last of section 3 C, the variation of $T_c$ with the *RE* dopant is opposite between the present $(Ba_{1-x}RE_x)Fe_2As_2$ epitaxial films and the $(Ca_{1-x}RE_x)Fe_2As_2$ single crystals; i.e., the maximum $T_c$ of the $(Ba_{1-x}RE_x)Fe_2As_2$ epitaxial films decreases in the order of *RE* = La, Ce, Pr and Nd, while that of the $(Ca_{1-x}RE_x)Fe_2As_2$ single crystals increases in this order [13–16]. The origins of this difference would be classified to an intrinsic effect and an extrinsic effect. To clarify the extrinsic effect originating from the different crystal structure parameters and crystalline quality between the thin films and the single crystals, we should use epitaxial films of *RE*-doped $CaFe_2As_2$ epitaxial films; however, unfortunately, epitaxial thin film growth of $CaFe_2As_2$ by PLD has not been attained [21]. Here, we like to discuss plausible intrinsic origins. One possible origin of this discrepancy may be the structural transition in $(Ca_{1-x}RE_x)Fe_2As_2$. It should be noted that $(Ca_{1-x}RE_x)Fe_2As_2$ transits easily to a collapsed tetragonal structure by replacing the Ca site with a smaller *RE* dopant, such as Pr and Nd [13]. In the thin film samples used here, the local structures of the $(Ba_{1-x}RE_x)Fe_2As_2$ films remained almost unchanged for different *RE*, implying that the film's electronic structures are not changed drastically by the *RE* doping. Therefore, it would be possible that the structural flexibility to a collapsed structure in $(Ca_{1-x}RE_x)Fe_2As_2$ contributes to the high $T_c$.

## 5. Conclusions

A non-equilibrium PLD process was used to stabilize nonmagnetic/magnetic *RE* (= La, Ce, Pr, and Nd) dopants in $BaFe_2As_2$ and their superconductivity were examined with respect to the 4*f* electrons of *RE* dopants. The indirect electron doping of each *RE* element produced superconductivity along with a suppression of the resistivity anomaly, which is associated with the magnetic/structural phase transitions. The La- and Ce-doping induced bulk superconductivity at a maximum $T_c^{onset}$ of 22.4 K at $x_{film}$ = 0.13 for La doping and 13.4 K at $x_{film}$ = 0.15 for Ce doping, respectively. In the cases of Pr- and Nd-doping, the superconducting transitions started from $T_c^{onset}$ = 6.2 K and 5.8 K, respectively, while zero resistivity and diamagnetism were not observed. The severe decease in $T_c^{onset}$ and SVFs upon *RE*-doping from non-magnetic La to magnetic Ce and Pr are attributed to magnetic electron pair breaking, as evidence of the magnetic electron scattering and the negative MR, with interaction between the localized 4*f* electrons and the Fe 3*d* conduction electrons.

**Acknowledgment**

This work was supported by the Japan Society for the Promotion of Science (JSPS), Japan, through the “Funding Program for World-Leading Innovative R&D on Science and Technology (FIRST Program)" and MEXT Element Strategy Initiative to Form Core Research Center.

**Figure captions**

**Figure 1.** High-resolution out-of-plane XRD patterns for films of $(Ba_{1-x}Ce_x)Fe_2As_2$ (a), $(Ba_{1-x}Pr_x)Fe_2As_2$ (b), $(Ba_{1-x}Nd_x)Fe_2As_2$ (c), and Sm-containing $BaFe_2As_2$ (d) with various $x_{film}$ values. The $x_{film}$ value is shown on the upper right of each panel. The vertical dashed lines indicate the 00*l* diffraction angles of the undoped $BaFe_2As_2$ phase. The asterisks show the 110 diffraction peaks from the Fe impurity phase.

**Figure 2.** $x_{film}$ dependence of structural properties of $(Ba_{1-x}RE_x)Fe_2As_2$ epitaxial films. (a) *a*-axis and *c*-axis lattice parameters and unit cell volume *V*, (b) FWHM values of rocking curves for 004 ($\Delta\omega$) and 200 ($\Delta\phi$) diffractions, and (c) FWHM values of 004 diffractions ($\Delta 2\theta$). The solid lines in the middle figure of (a) are indicators of changes in the *c*-axis lattice parameters. Open pentagon symbols and dotted lines show the results for undoped $BaFe_2As_2$ films. The diamonds, circles, squares, and triangles indicate values for La-, Ce-, Pr-, and Nd-doped films, respectively.

**Figure 3.** $\rho$–$T$ curves in the temperature range from 300 to 2 K for (a) $(Ba_{1-x}Ce_x)Fe_2As_2$ ($x_{film}$=0–0.29), (b) $(Ba_{1-x}Pr_x)Fe_2As_2$ ($x_{film}$=0.06–0.18), and (c) $(Ba_{1-x}Nd_x)Fe_2As_2$ ($x_{film}$=0.07–0.13) epitaxial films, respectively. The doping concentration $x_{film}$ is indicated on the upper left of each panel. The inset figures show magnified views around $T_c$. The arrows and triangles indicate the positions of $T_{anom}$ and $T_{min}$, respectively. The black curved lines are fitting results from conventional power-law behavior, $\rho = \rho_0 + AT^n$ ($\rho_0$: residual resistivity), in the higher-temperature regions.

**Figure 4.** $\rho$–$T$ curves under magnetic fields of 0–9 T for optimally doped

$(Ba_{0.85}Ce_{0.15})Fe_2As_2$ (a), $(Ba_{0.82}Pr_{0.18})Fe_2As_2$ (b), and $(Ba_{0.87}Nd_{0.13})Fe_2As_2$ (c) epitaxial films. The inset figures show $\chi$–$T$ curves under a magnetic field of 5 G.

**Figure 5.** Electronic phase diagrams of $(Ba_{1-x}RE_x)Fe_2As_2$ epitaxial films. The diamonds, circles, squares, and triangles represent phase diagrams for the epitaxial films of $(Ba_{1-x}La_x)Fe_2As_2$ [18], $(Ba_{1-x}Ce_x)Fe_2As_2$, $(Ba_{1-x}Pr_x)Fe_2As_2$, and $(Ba_{1-x}Nd_x)Fe_2As_2$, respectively. $T_{anom}$ and $T_c^{onset}$ are indicated by the closed and open symbols, respectively. The vertical dashed/dotted lines indicate the solubility limits for the dopants. The dashed curves show phenomenological fits to $T_{anom}$ as a function of $x_{film}$, and the bottom vertical arrows indicate the critical $x$ points extrapolated to $T_{anom}$=0. The inset figure represents the $c$-axis lattice parameter dependences of $T_{anom}$ and $T_c^{onset}$.

**Figure 6.** Field dependence of magnetoresistance MR = $[\rho(\mu_0 H)-\rho(0)]/\rho(0)\times 100$ at 30 K for epitaxial films of (a) $(Ba_{1-x}La_x)Fe_2As_2$ with $x_{film}$=0–0.44, (b) $(Ba_{1-x}Ce_x)Fe_2As_2$ with $x_{film}$=0.06–0.29, and (c) $(Ba_{1-x}Pr_x)Fe_2As_2$ with $x_{film}$=0.06–0.18. The $x_{film}$ values for each MR are indicated in the figures. The solid red lines in (a) represent the fitting results with the $MR_{SDW}$ model, and those in (b) and (c) show the fitting results with a $MR_{SDW}-\alpha H^2$ model. The dotted lines in (b) and (c) indicate the fitting results with a $MR_{SDW}-\beta \log H$+const. and the arrows represent the transition point from a $-\alpha H^2$ dependence to a $-\alpha \log H$ dependence.

**Table. I.** Summary of superconducting and structural properties of optimally doped epitaxial films of $(Ba_{0.87}La_{0.13})Fe_2As_2$, $(Ba_{0.85}Ce_{0.15})Fe_2As_2$, and $(Ba_{0.82}Pr_{0.18})Fe_2As_2$.

**Table. II.** Summary of the fitting parameters using the model [39] for the magnetoresistance of $(Ba_{1-x}La_x)Fe_2As_2$ epitaxial films. Errors are indicated in the values in the parentheses.

# Figures

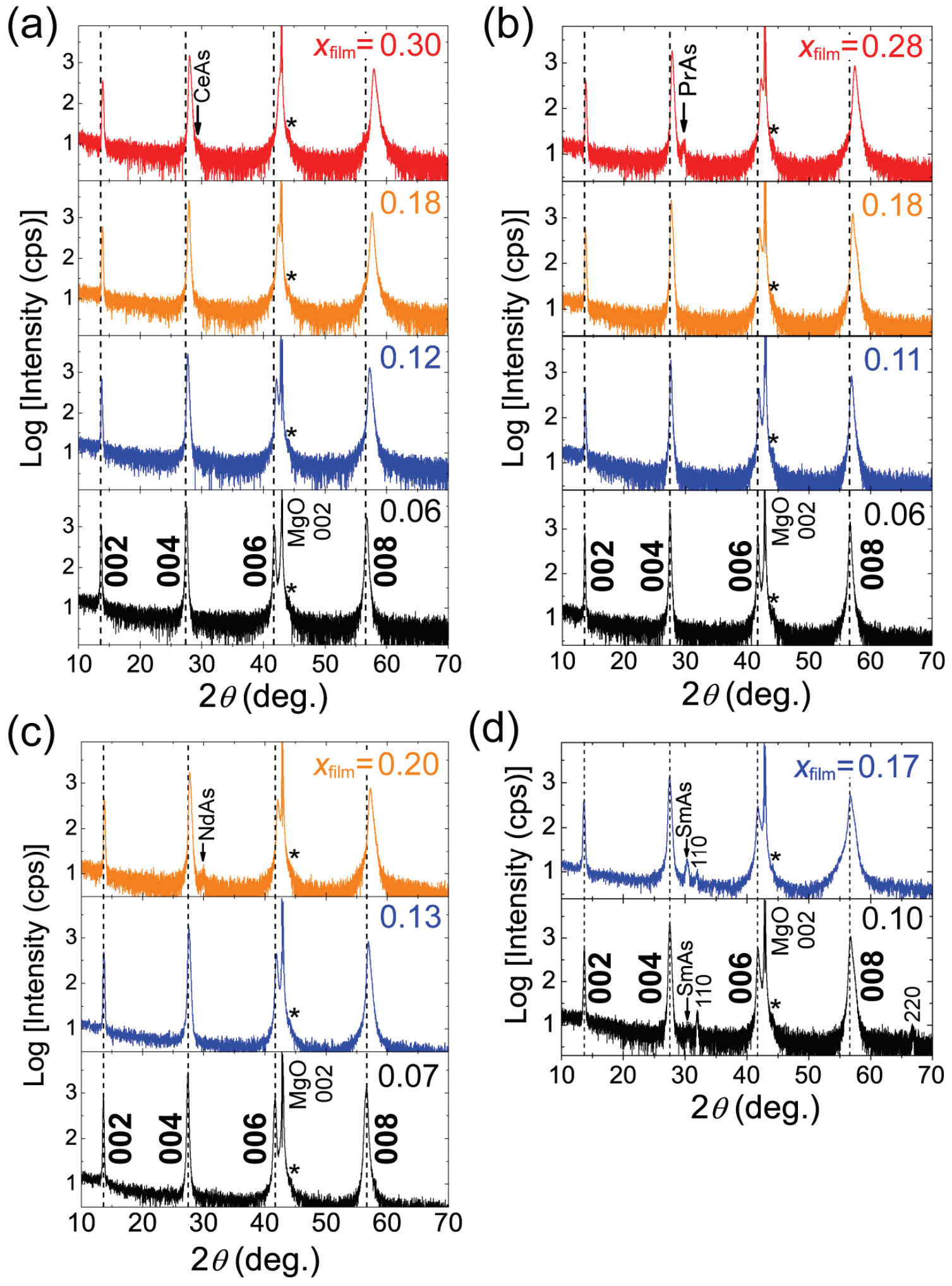

Figure 1.

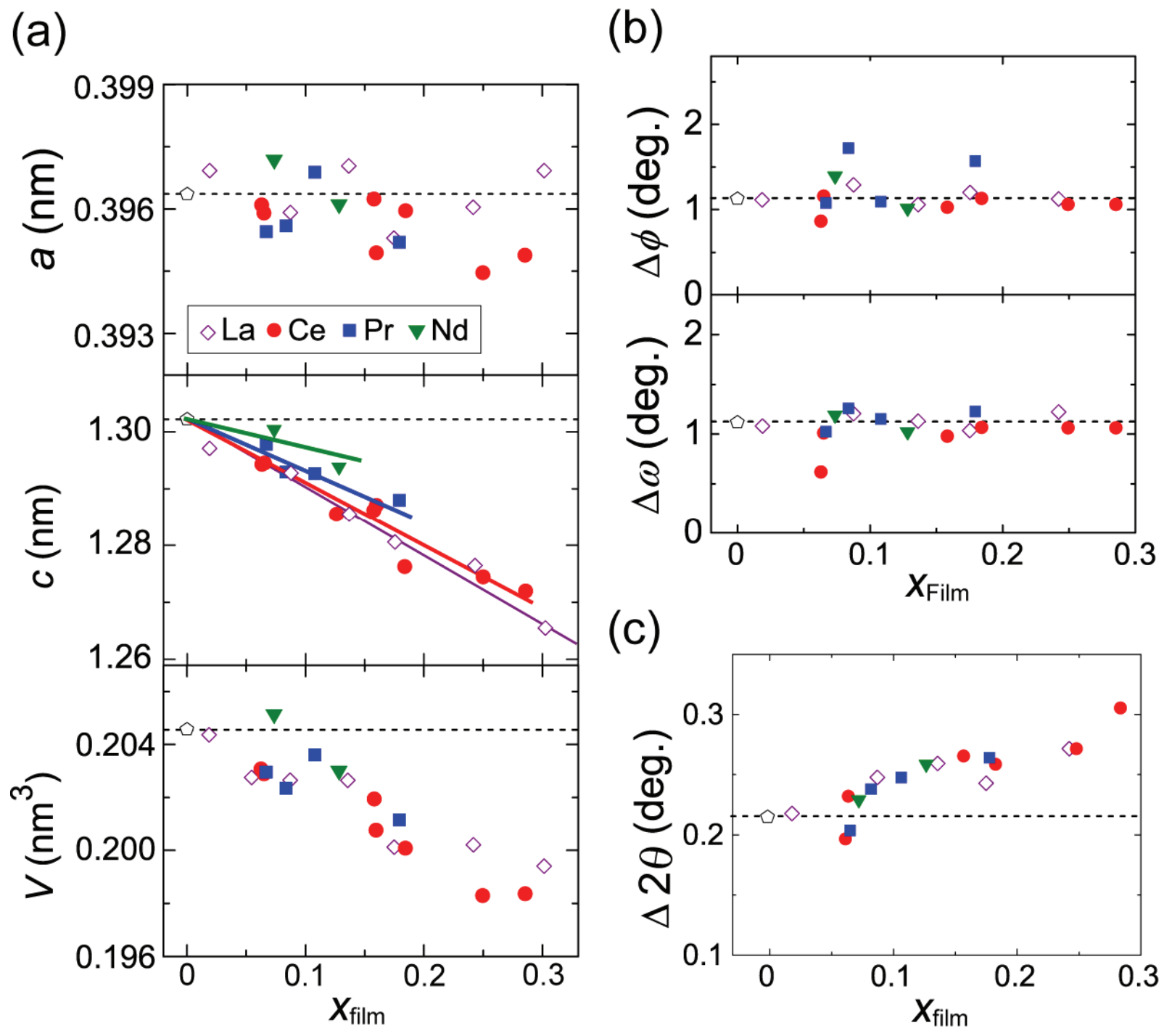

Figure 2.

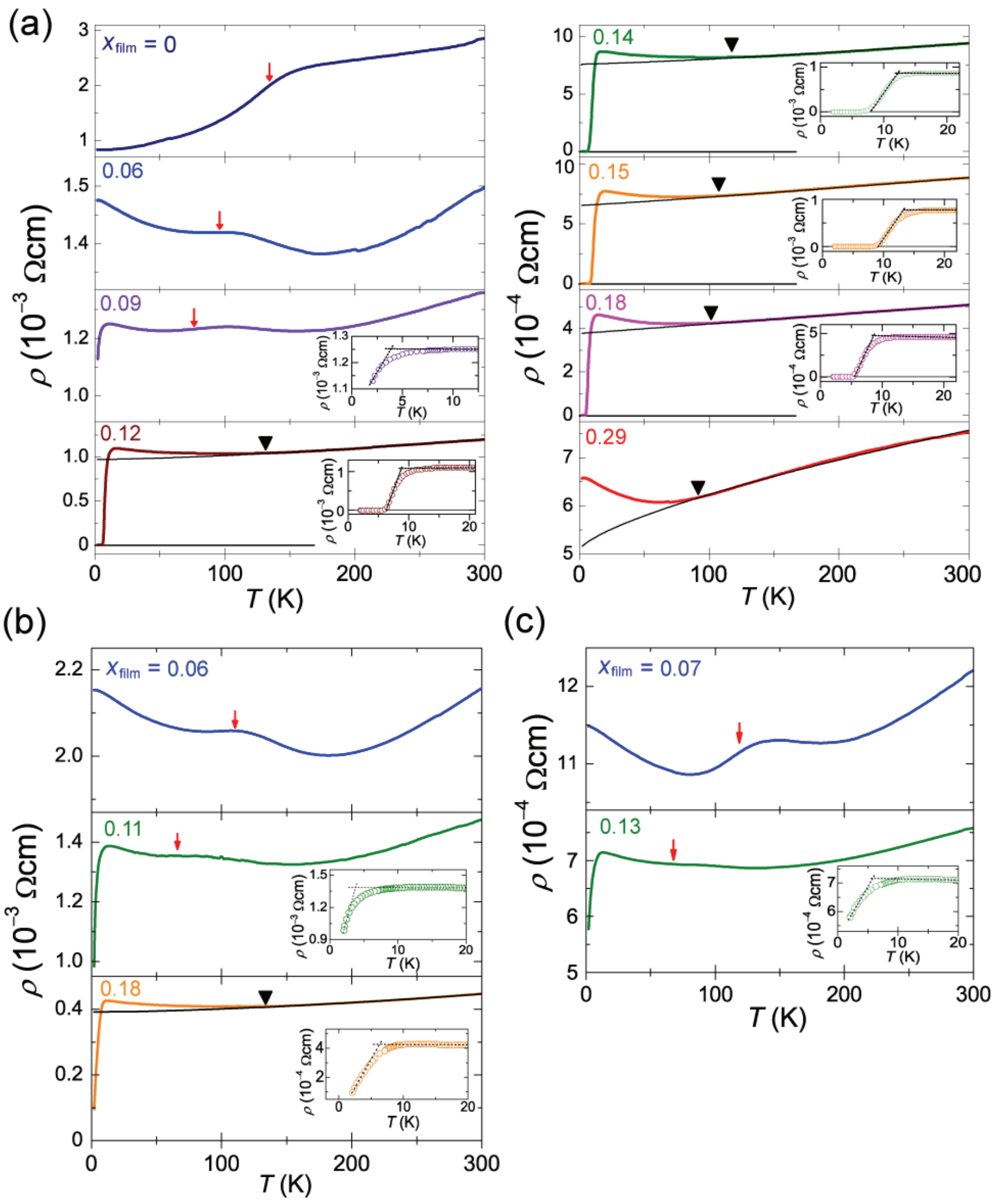

Figure 3.

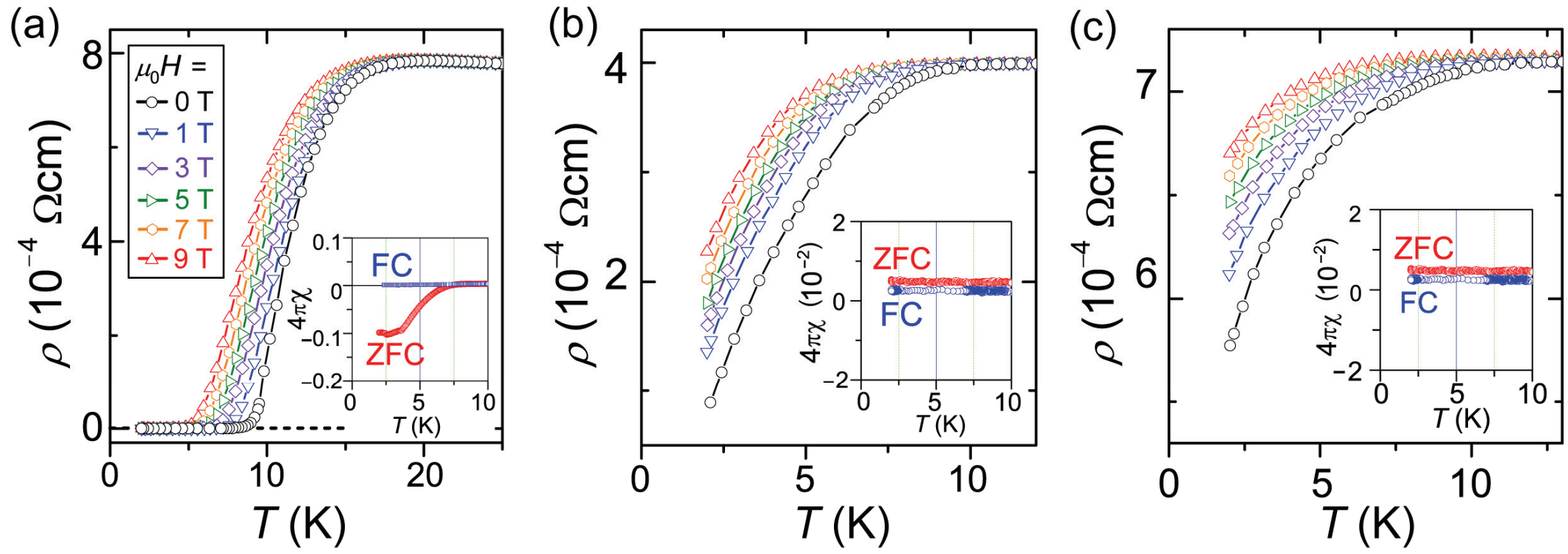

Figure 4.

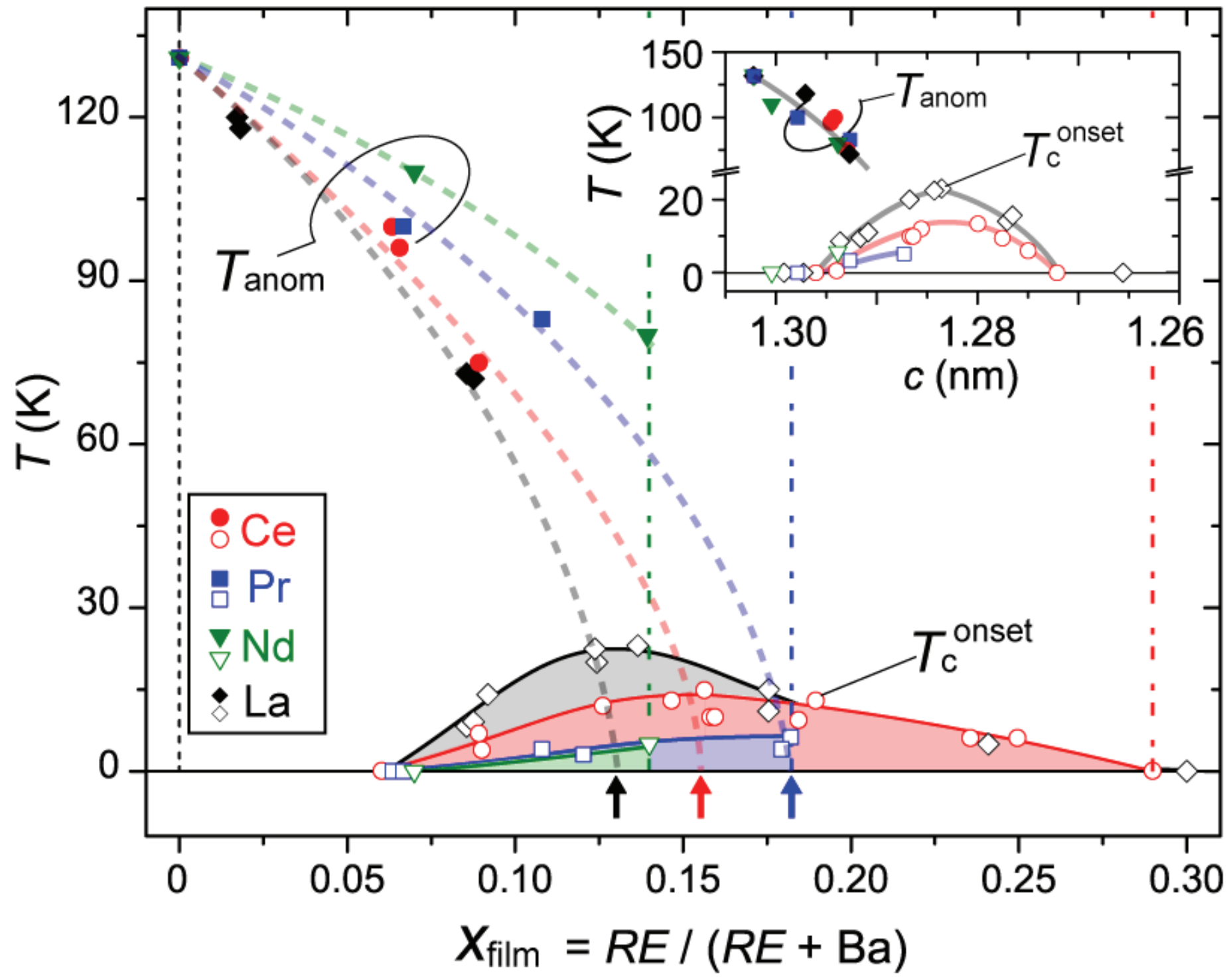

Figure 5.

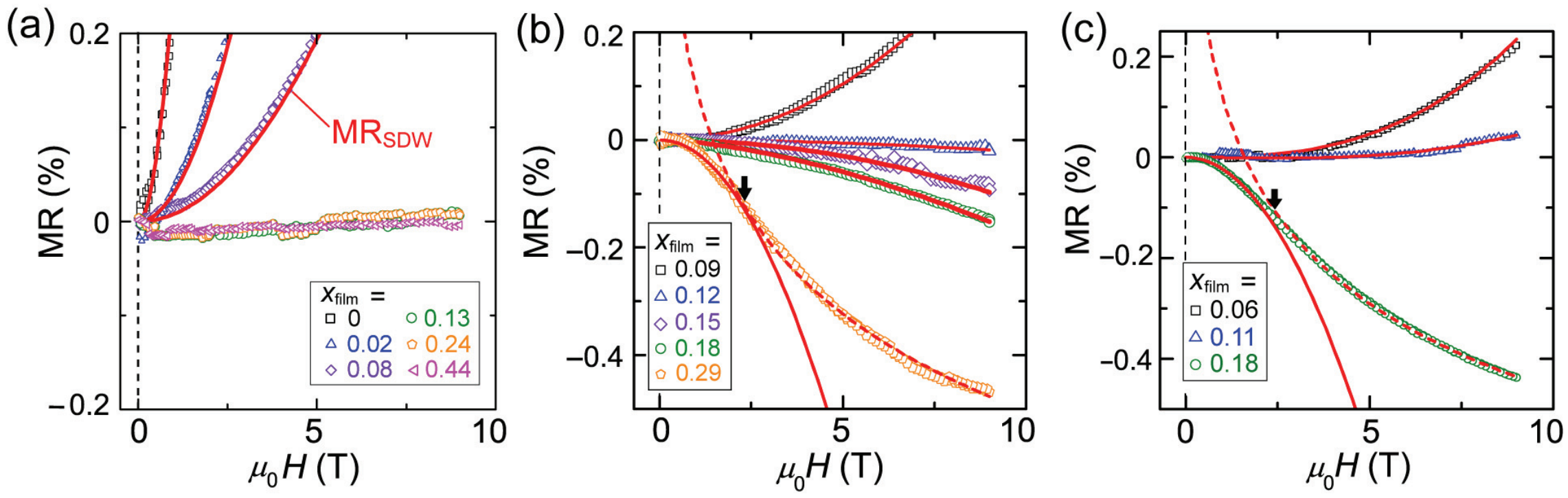

Figure 6.

Table 1.

| | Optimal $x_{film}$ | $T_c^{onset}$ (K) | $c$ (nm) | $a$ (nm) | $z_{As}$ | $h_{As}$ (nm) | α (deg.) |
|---|---|---|---|---|---|---|---|
| $(Ba_{1-x}La_x)Fe_2As_2$ | 0.13 | 22.4 | 1.2872 | 0.3970 | 0.352 | 0.13129 | 113.04 |
| $(Ba_{1-x}Ce_x)Fe_2As_2$ | 0.15 | 13.4 | 1.2866 | 0.3962 | 0.352 | 0.13123 | 112.95 |
| $(Ba_{1-x}Pr_x)Fe_2As_2$ | 0.18 | 6.2 | 1.2879 | 0.3952 | 0.352 | 0.13136 | 112.76 |

Table 2.

| | $T$ | $\beta$ | $\gamma$ | $\mu_0$ ($m^2$/Vs) |
|---|---|---|---|---|
| $x_{film}$ = 0 in $(Ba_{1-x}La_x)Fe_2As_2$ film | 30 | 34.0 (±0.1) | 1.75 (±0.05) | 0.151 (±0.001) |
| $x_{film}$ = 0.02 in $(Ba_{1-x}La_x)Fe_2As_2$ film | 30 | 79.0 (±0.3) | 1.25 (±0.06) | 0.0740 (±0.001) |
| $x_{film}$ = 0.08 in $(Ba_{1-x}La_x)Fe_2As_2$ film | 30 | 98.7 (±0.8) | 1.52 (±0.06) | 0.0430 (±0.0004) |
| $BaFe_2As_2$ single crystal (Ref. 39) | 50 | 44.5 | 1.14 | 0.16 |

Supplementary Information for

**"Magnetic scattering and electron pair breaking by rare-earth-ion substitution in $BaFe_2As_2$ epitaxial films"**

Takayoshi Katase[1,4], Hidenori Hiramatsu[2,3], Toshio Kamiya[2,3] and Hideo Hosono[1,2,3,5]

[1] Frontier Research Center, Tokyo Institute of Technology, S2-6F East, 4259 Nagatsuta-cho, Midori-ku, Yokohama 226-8503, Japan

[2] Materials and Structures Laboratory, Tokyo Institute of Technology, 4259 Nagatsuta-cho, Midori-ku, Yokohama 226-8503, Japan

[3] Materials Research Center for Element Strategy, Tokyo Institute of Technology, 4259 Nagatsuta-cho, Midori-ku, Yokohama 226-8503, Japan

[4] Present address: Research Institute for Electronic Science, Hokkaido University, Sapporo 001-0020, Japan

[5] Corresponding author. E-mail: hosono@msl.titech.ac.jp

## I. Synthesis of bulk polycrystal samples of rare earth (*RE*) containing $BaFe_2As_2$

Bulk polycrystal samples of *RE*- (= Ce, Pr, Nd, and Sm) containing $BaFe_2As_2$ were synthesized for the PLD targets by two-step solid-state reactions. All preparation procedures except for the annealing process were carried out in an Ar-filled glove box (the $O_2$ impurity concentration was < 1 ppm and the dew point was < –100°C). A mixture of fine pieces of Ba and *RE* metals along with the powders of Fe and As were mixed in a stoichiometric atomic ratio of Ba : *RE* : Fe : As = 1–*x* : *x* : 2 : 2 and sealed in a stainless-steel tube under a pure Ar atmosphere, followed by a reaction at 700°C for 10 h. The resulting powders were grounded thoroughly and pressed into pellets. The pellets were placed in stainless-steel tubes and heated at 900°C for 16 h. It is important to carefully avoid contamination with atmospheric impurities such as oxygen- and water-related molecules in the PLD targets in order to obtain the high-purity $(Ba_{1-x}RE_x)Fe_2As_2$ films; i.e., we could not succeed in the *RE* doping at the early stages of this study when a small amount of a *RE*FeAsO impurity phase was observed in the PLD targets. Further, it should be noted that the *RE*FeAsO phase was always produced when we synthesized $(Ba_{1-x}RE_x)Fe_2As_2$ using an agate mortar and evacuated silica glass tubes similar to the case to synthesize $Ba(Fe,Co)_2As_2$ in ref. S1. We, therefore, used a fully-dried silica-glass mortar to mix the powders and then sealed the pressed pellets into stainless tubes before the reactions.

Powder X-ray diffraction (XRD; anode radiation: CuKα, D8 ADVANCE-TXS, Bruker AXS) was measured at room temperature for the resulting bulk polycrystals. Powder XRD patterns of the $(Ba_{1-x}RE_x)Fe_2As_2$ bulk polycrystal samples with nominal $x$ = 0.1 are shown in Fig. S1(a), and the nominal $x$ dependence of the lattice parameters in Fig. S1(b). No peak shift was observed in the diffraction angles of the $(Ba_{1-x}RE_x)Fe_2As_2$ samples from the calculated peak positions of undoped $BaFe_2As_2$. The lattice shrinkage, which is considered to originate from the substitution of Ba with the *RE* ions having smaller ion radii, was not observed, even though $x$ was varied up to 0.25. Simultaneously, the 200 diffractions for *RE*As impurities were observed distinctly for all the cases as indicated by the arrows in Fig. S1(a). These phase-separated bulk polycrystals (i.e., $BaFe_2As_2$ + *RE*As) were used as the PLD targets.

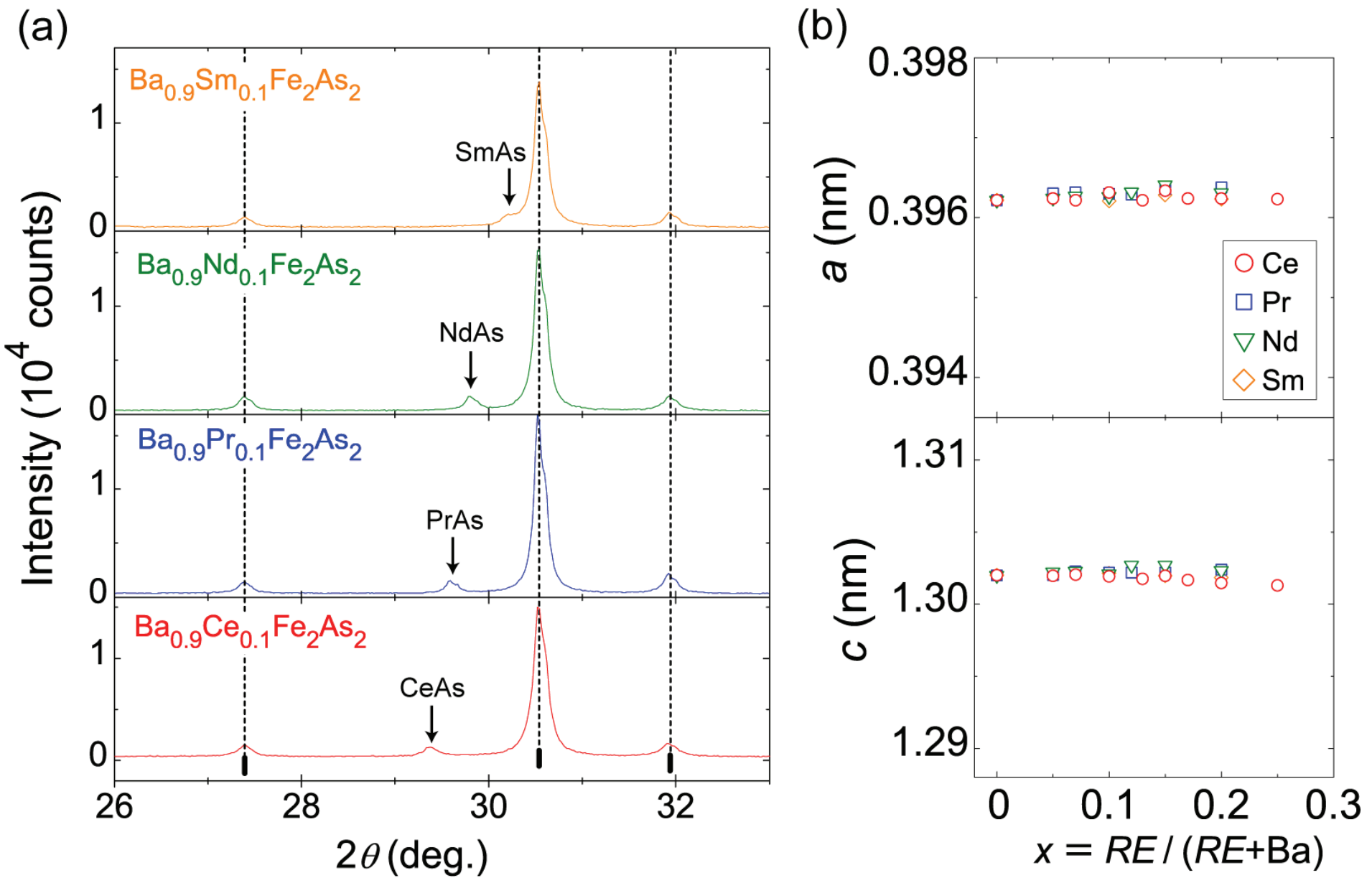

**Figure S1.** (a) Powder XRD patterns of $(Ba_{1-x}RE_x)Fe_2As_2$ bulk polycrystal samples with $x$ = 0.1. The nominal chemical composition of the bulk crystal is indicated on upper left in each pattern. The vertical dotted lines indicate the diffraction angles of undoped $BaFe_2As_2$. (b) The nominal $x$ dependence of the *a*- and *c*-axis lattice parameters for the $(Ba_{1-x}RE_x)Fe_2As_2$ bulk polycrystal samples. The circle, square, triangle, and diamond symbols correspond to the results of the Ce, Pr, Nd, and Sm containing $BaFe_2As_2$, respectively.

## II. Chemical composition analysis of $(Ba_{1-x}RE_x)Fe_2As_2$ epitaxial films

Figure S2(a) shows the *RE* chemical composition mapping images for the optimally doped epitaxial films of $(Ba_{0.85}Ce_{0.15})Fe_2As_2$, $(Ba_{0.82}Pr_{0.18})Fe_2As_2$, and $(Ba_{0.87}Nd_{0.13})Fe_2As_2$ examined with an electron-probe microanalyzer (EPMA). The EPMA mapping images ensured homogeneous *RE* distribution. Figure S2(b) presents the measured doping concentration $x_{Film}$ in the $(Ba_{1-x}RE_x)Fe_2As_2$ epitaxial films measured by EPMA as a function of the nominal $x$. The good linearity in the relationship between the measured $x$ and the nominal $x$ demonstrates that the chemical compositions of the PLD targets were transferred linearly to the films. The solubility limit lines, which shifted to lower $x_{film}$ as *RE* changes in the order of Ce, Pr, and Nd, were determined as the chemical composition region where an impurity phase was not detected by the XRD measurements in Fig. 1.

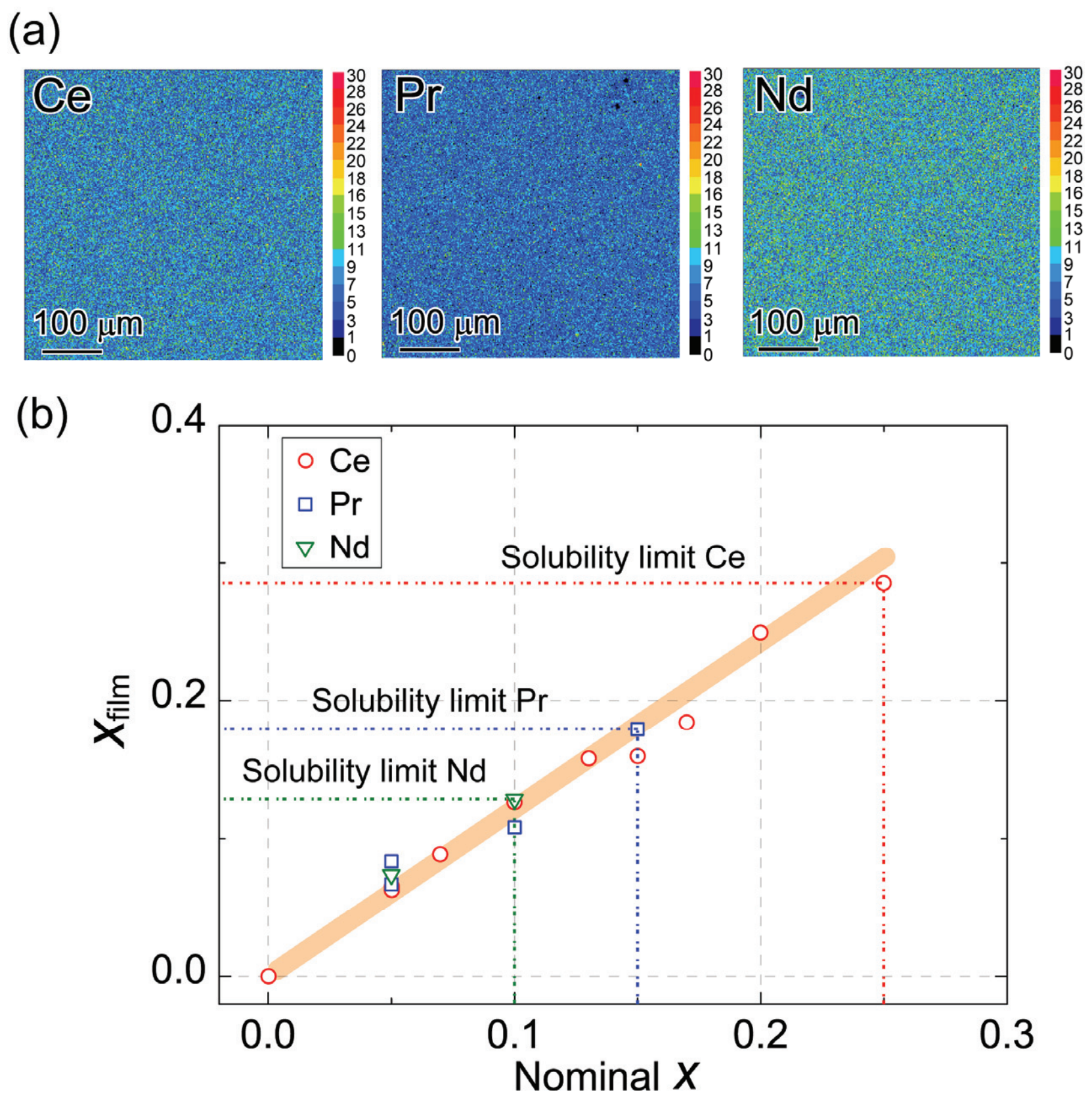

**Figure S2** (a) EPMA mapping images of *RE* elements in epitaxial films of $(Ba_{0.85}Ce_{0.15})Fe_2As_2$ (left), $(Ba_{0.82}Pr_{0.18})Fe_2As_2$ (middle), and $(Ba_{0.87}Nd_{0.13})Fe_2As_2$ (right). The horizontal bars at bottom left in the maps indicate a 100 μm scale, and the numbers on right of the color scale bars indicate the measured signal intensity in counts per pixel. (b) Relationship between the $x_{film}$ and the nominal $x$ measured by EPMA. The circle, square, and triangle symbols indicate Ce-, Pr-, and Nd-doping results, respectively. The dotted lines represent the solubility limits.

## III. Temperature derivative of resistivity

Figure S3 shows the temperature derivatives of resistivities (d$\rho$/d$T$) for the films in Fig. 3 [i.e., $(Ba_{1-x}Ce_x)Fe_2As_2$ with $x_{film}$ = 0–0.12, $(Ba_{1-x}Pr_x)Fe_2As_2$ with $x_{film}$ = 0–0.18, and $(Ba_{1-x}Nd_x)Fe_2As_2$ with $x_{film}$ = 0–0.13]. The resistivity anomaly temperature ($T_{anom}$), defined as the peak position in the d$\rho$/d$T$ curves, shifts to lower $T$ with increasing the doping concentrations ($x_{film}$).

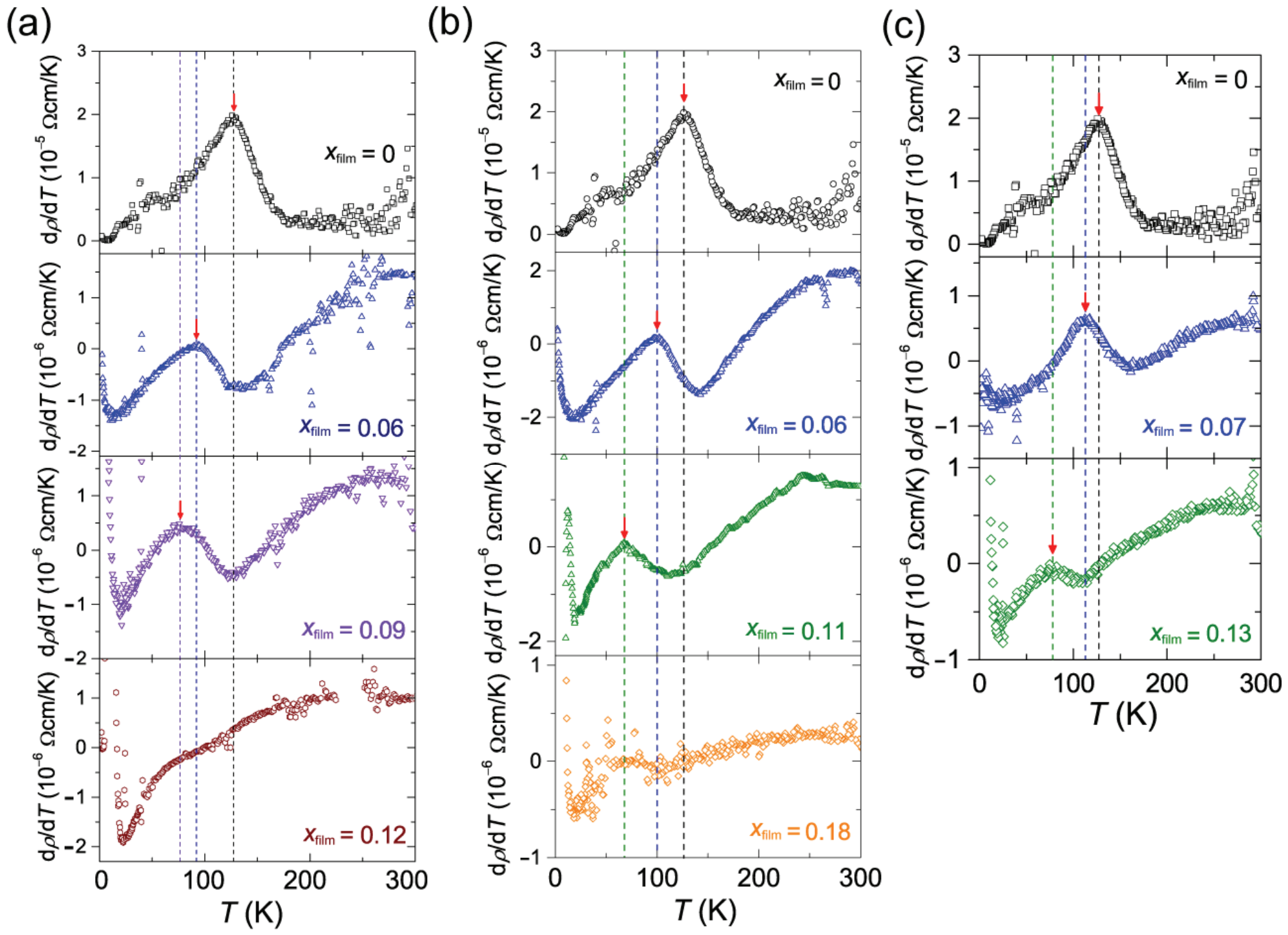

**Figure S3** d$\rho$/d$T$ curves of (a) $(Ba_{1-x}Ce_x)Fe_2As_2$ with $x_{film}$ = 0–0.12, (b) $(Ba_{1-x}Pr_x)Fe_2As_2$ with $x_{film}$ = 0–0.18, and (c) $(Ba_{1-x}Nd_x)Fe_2As_2$ with $x_{film}$ = 0–0.13. The doping concentration $x_{film}$ is indicated on right-bottom of each figure. The positions of $T_{anom}$ are indicated by the red vertical arrows.

## IV. Field dependence of magnetoresistance

Figure S4 shows the magnetoresistance (MR) of $(Ba_{0.72}Ce_{0.28})Fe_2As_2$ epitaxial film measured at (a) 30K, (b)10 K, and (c) 2 K. The MR were measured under magnetic field parallel to the *c*-axis and the *ab*-plane, respectively. With decreasing the temperature, the negative MR steeply increases and the transition magnetic field from the $-\alpha H^2$ dependence to the $-\alpha\log H$ dependence shifts to a lower $H$. In addition, the negative MR with *H*//*ab* is much smaller than that with *H*//*c* for all the temperatures. This result indicates that the spin ordering of Ce should be antiferromagnetic parallel to the *c*-axis and ferromagnetic parallel to the *ab*-plane.

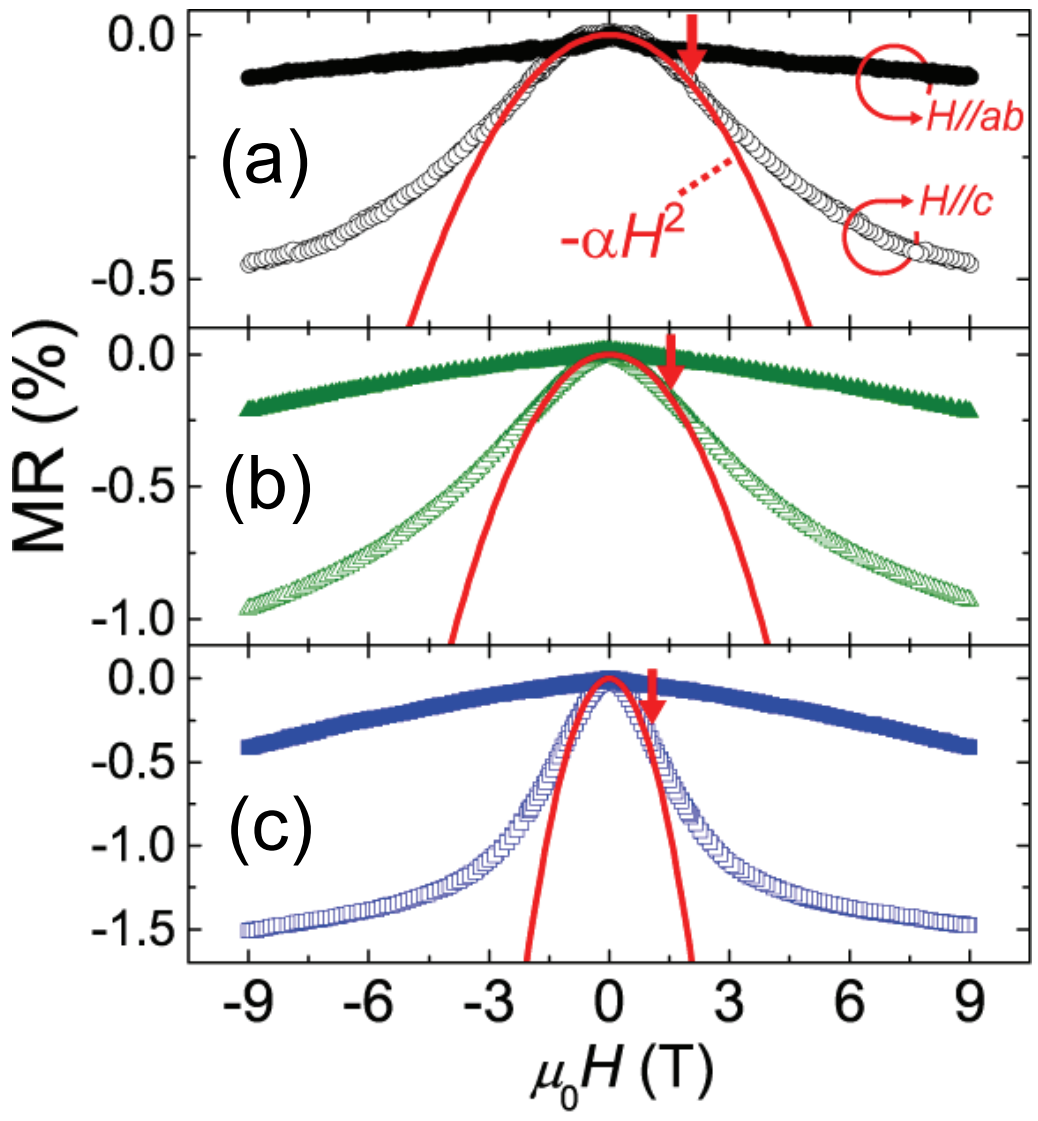

**Figure S4** MR = [$\rho(\mu_0H)-\rho(0)]/\rho(0)\times100$] measured at (a) 30 K, (b) 10 K, and (c) 2 K for $(Ba_{0.72}Ce_{0.28})Fe_2As_2$ epitaxial film. Open and closed symbols indicate MR under fields ($H$) parallel to $c$-axis and to the $ab$-plane, respectively. The electric current for MR was applied along the $ab$-plane and perpendicular to both of the magnetic field directions. The solid red lines represent the fitting results with the $-\alpha H^2$ behavior. The red vertical arrows indicate the transition magnetic fields from the $-\alpha H^2$ dependence to the $-\alpha\log H$ dependence.